\documentclass[12pt]{article}
\usepackage{latexsym}
\usepackage{amsmath}
\usepackage{amsfonts}
\usepackage{amssymb}
\usepackage{graphicx}
\usepackage{slashbox}
\usepackage{color}
 
\newcommand{\mR}{{\mathbb R}}

\title{On dynamical realizations of l-conformal Galilei groups}
\author{K. Andrzejewski\thanks{e-mail: k-andrzejewski@uni.lodz.pl},
J. Gonera, P. Kosi\'nski, P. Ma\'slanka\thanks{e-mail: pmaslan@uni.lodz.pl}\\
\small Department of Theoretical Physics and Computer Science, \\
\small University of \L\'od\'z,\\
\small Pomorska 149/153, 90-236 {\L}\'od\'z, Poland
}
\date{}
\begin{document}
\maketitle
\begin{abstract}
We consider the dynamics invariant under the action of $l-$conformal Galilei group using the method of nonlinear realizations.
 We find that by an appropriate choice of the coset space
parametrization one can achieve the complete decoupling of the equations of motion. The Lagrangian and Hamiltonian are constructed. 
The results are compared with those obtained 
by Galajinsky and Masterov [Nucl. Phys. {\bf B866}, (2013), 212].
\end{abstract}
\section{Introduction}
It is well known that the Galilei algebra posses the whole family of conformal extensions indexed by a positive integer or half-integer $l$\ \cite{b1}. The $l=1/2$\ member was
 discovered in nineteenth century \cite{b2} and rediscovered in twentieth century as the maximal symmetry algebra of free motion in quantum mechanics \cite{b3}. Recently the higher $l-$ 
conformal algebras have also attracted  some attention \cite{b4}.

\par
The basic question concerning any symmetry group is the construction and, if possible, classification of admissible invariant dynamics. In the case of the centrally extended $l-$conformal algebras/groups the problem can be solved \cite{b5}, \cite{b6}  using the orbit method \cite{b7}, at least under the assumption that the symmetry group acts transitively.
\par
However, the central extension is admissible only for $l$\ half-integer or $(1+2)$-dimensional spacetime \cite{b8}. If the central extension doesn't exist or the central charge vanishes
the situation is more complicated. Although the orbit method is still applicable, the full classification of coadjoint orbits is rather involved.
\par
Alternatively, in order to construct invariant dynamical systems the theory of nonlinear group realizations \cite{b9},\cite{b10} can be used.  \ To this end one computes the
 relevant Cartan forms and imposes the invariant constraints on them (this procedure is sometimes known under the name of inverse Higgs phenomenon \cite{b11}).
It can be shown that, in the case when the coset manifold entering the construction is isomorphic to some coadjoint orbit, the technique of nonlinear realizations is equivalent to the 
orbit method \cite{b12}. However, it is applicable in the case of more general manifolds. The main trouble here is that, contrary to the orbit method, it doesn't automatically
lead to Hamiltonian form of dynamics.
\par
The technique of nonlinear realizations has been successfully applied to the $l=1$\ conformal Galilei algebra \cite{b13}. The results obtained there were generalized to 
arbitrary $l$\ in Refs. \cite{b14} and \cite{b15}. The invariant equations appear to describe decoupled $SL(2,\mR)$\ conformal mode as well as the dynamics of additional 
coordinates related to the generators lying outside the $so(3)\oplus sl(2,\mR)$\ subalgebra. The latter is shown to be related to higher derivative free dynamics via
 generalized Niederer's transformation \cite{b16}. It has been also shown in Ref. \cite{b14} that one can construct invariant dynamics in terms of second order
 differential equations. However, the problem of their Hamiltonian description remains open. 
\par
In the present paper we analyse further the $l$-conformal Galilean dynamics within the framework of nonlinear realizations. Let us remind the essential points of the method.
Given a Lie group $G$\ one selects a subgroup $S\subset G$\ (the "stability" subgroup ). The main assumption is that under the adjoint action of $S$\ the Lie algebra $\mathcal G$
of $G$\ decomposes into the direct  sum of representation spanned by the subalgebra $\mathcal S$\ and the one spanned by the remaining generators of $\mathcal G$. All dynamical variables 
are classified into two types: the preferred (or Goldstone) variables and the adjoint ones. The former parametrize the coset manifold $G/S$; the latter transform under $G$\
 according to some representation  of the subgroup $S$, with group parameters depending both on initial group element and the preferred variables. The Goldstone variables enter the
 Lagrangian only through covariant derivatives which again transform according to some representation of $S$. Therefore, in order to construct the Lagrangian invariant under
 nonlinear action of $G$\ it is sufficient to construct the Lagrangian invariant under the linear action of $S$.  
\par
The method of nonlinear realizations can be applied to the case of spacetime symmetries \cite{b17}. The additional complication here is that some variables play a double  
role of Goldstone modes and spacetime arguments of the remaining ones. This results in some modifications of covariant derivatives (and, in quantum field theory, volume elements)
 which appear to be the ratios of certain Cartan forms.
\par
Below we use the formalism of nonlinear realizations to analyze the dynamics invariant under $l$-conformal Galilean transformations. The approach based on inverse Higgs phenomenon 
makes use of invariant constraints imposed on Cartan forms. There are basically two kinds of such constraints. First, there are constraints that serve to eliminate some variables
in favour of the others. The remaining ones define the dynamics invariant under the group action. As far as the reduction of the number of dynamical variables  is concerned one
 can proceed in a slightly different way, namely by enlarging the dimension of the stability subgroup thus reducing the number of Goldstone variables. To illustrate this we
 consider in Sec. II, as a warming-up exercise, the case of conformal mechanics. The stability subgroup is chosen to consist of dilatations while the Goldstone variables are
 time together with one generalized coordinate. It is then easy to build an invariant Lagrangian. Moreover, it appears that the result is equivalent to the one obtained in 
Ref. \cite{b18} where the action of the group on whole group manifold is the starting point and in Ref. \cite{b12} where, in turn, the stability subgroup is generated by $H+K$.  
\par 
In Sec III we consider the nonlinear realizations of $l$-conformal Galilei group linearizing on rotations and dilatations. The equations of motion are obtained by imposing 
constraints on Cartan forms related to the generators outside  $so(3)\oplus sl(2,\mR)$. By choosing a modified parametrization of coset manifold we reduce these equations 
to very simple form describing higher-derivative free motion. The conformal mode decouples completely and becomes a kind of internal
 variable. It is easy to construct the 
first-order invariant Lagrangian which describes a constrained system in the sense of Dirac. Standard technique yields then the Hamiltonian formalism described in sec.IV.
\par
Sec.V is devoted to the short comparison of our approach with that of Galajinsky and Masterov \cite{b14}. We also mention there the results obtained previously for $l$\
half-integer and compare them with the present findings.
\par
Sec. VI is devoted to some conclusions.

\section{Conformal mechanics} 
The prototype of all conformal groups is the one acting in $1+0$-dimensional spacetime, locally isomorphic to $SL(2,\mR)$; in fact it is $SO(2,1)\simeq SL(2,\mR)/\{I,-I\}$.
The corresponding Lie algebra is spanned by $H$ (time translation), $D$ (dilatation) and $K$ (special conformal transformation) obeying

\begin{equation}
\label{e1}
\begin{split}
[D,H]&=iH,  \\
[D,K]&=-iK,  \\
[K,H]&=2iD. 
\end{split}
\end{equation}
The simplest dynamical system invariant under the action of $SL(2,\mR)$\ group was constructed in  the first Ref. \cite{b3} and in \cite{b19}.
 Its geometrical structure was investigated
 in the elegant paper by Ivanov et al. \cite{b18} where the method of nonlinear realizations was used. Further analysis, with the 
help of orbit method \cite{b7}, 
is given in Ref. \cite{b12}. It appears that the common basis for both methods is the symmetric Cartan decomposition of $sl(2,\mR)$\ 
algebra with respect to the
 compact generator $H+K$.
\par
Let us consider the symmetric Cartan decomposition based on $D$ as the stability subgroup generator. The coset space is parametrized 
as follows    
\begin{equation}
\label{e2}
\begin{split}
w=e^{itH}e^{izK}
\end{split}
\end{equation}
and the action of $SL(2,\mR)$\ is defined by
\begin{equation}
\label{e3}
\begin{split}
ge^{itH}e^{izK}=e^{it'H}e^{iz'K}e^{iu'D}
\end{split}
\end{equation}
According to the terminology of Ref. \cite{b9} $t$ and $z$ are preferred variables. The action of
$
 g=\left(
\begin{array}{cc}
\alpha &\beta\\
\gamma &\delta
\end{array}
\right)\in SL(2,\mR) $
   is easily found by using the representation spanned by
\begin{equation}
\label{e4}
\begin{split}
H=-i\sigma _+, \quad K=i\sigma _-, \quad D=-\frac{i}{2}\sigma _3 .
\end{split}
\end{equation}
It reads
\begin{align}
\label{e5}
&t'=\frac{\alpha t+\beta }{\gamma t+\delta },\nonumber\\
&z'=(\gamma t+\delta )^2z-\gamma (\gamma t+\delta ),\\
&u'=-2\ln (\gamma t+\delta ).\nonumber
\end{align}
The adjoint variables $\eta $\ transform according to
\begin{equation}
\label{e6}
\begin{split}
\eta '=e^{u'd_\eta }\eta =(\gamma t+\delta )^{-2d_\eta }\eta 
\end{split}
\end{equation}
$d_\eta $\ being the dilatational dimension of $\eta $.
The Cartan forms $w^{-1}dw\equiv i(\omega _HH+\omega _KK+\omega _DD)$\ read
\begin{align}
\label{e7}
&\omega _H=dt\nonumber,\\
&\omega _K=dz+z^2dt,\\
&\omega _D=-2zdt,\nonumber
\end{align}
 and transform according to the rules 
\begin{align}
\label{e8}
&\omega _H'=e^{u'}\omega _H=(\gamma t+\delta )^{-2}\omega _H,\nonumber\\
&\omega _K'=e^{-u'}\omega _K=(\gamma t+\delta )^2\omega _K,\\
&\omega _D'=\omega _D-du'=\omega _D+\frac{2\gamma dt}{\gamma t+\delta }.\nonumber
\end{align}  
The covariant derivative of $z$\ is defined as the ratio of Cartan forms \cite{b17}  
\begin{equation}
\label{e9}
\begin{split}
\nabla z=\frac{\omega _K}{\omega _H}=\dot{z} +z^2 ;
\end{split}
\end{equation}  
 one can easily obtains 
\begin{equation}
\label{e10}
\begin{split}
\nabla z'=e^{-2u'}\nabla z=(\gamma t+\delta )^4\nabla z .
\end{split}
\end{equation}  
 In order to construct the invariant dynamics it is sufficient to find the action integral invariant under the dilatation subgroup. It reads 
 \begin{equation}
\label{e11}
\begin{split}
S=\sigma \int {\omega _H\sqrt{\nabla z}}=\sigma \int {dt\sqrt{\dot{z}+z^2}} .
\end{split}
\end{equation} 
where $\sigma $\ is an arbitrary constant.\\
 Eq.(\ref{e11}) leads to the following equation of motion   
 \begin{equation}
\label{e12}
\begin{split}
\ddot{z} +6z\dot{z} +4z^3=0.
\end{split}
\end{equation}
 It is invariant under the action of the conformal group given by eq. (\ref{e5}).

Note in passing that the eq. (\ref{e12})  describes the whole family of standard conformal models. To see this we make a substitution  suggested by the constraint imposed on the Cartan form related to dilatation generator \cite{b14,b18}

 \begin{equation}
\label{e13}
\begin{split}
z=\frac{\dot{\rho }}{\rho }  .
\end{split}
\end{equation} 
Then eq. (\ref{e12}) yields
\begin{equation}
\label{e14}
\begin{split}
\frac{d}{dt}(\ddot{\rho }\rho ^3)=0  ,
\end{split}
\end{equation}
or
\begin{equation}
\label{e15}
\begin{split}
\ddot{\rho }=\frac{\gamma ^2}{\rho ^3} .
\end{split}
\end{equation}
We  have chosen the integration constant $\gamma^2>0$ because the conformal mechanics with repulsion is consistent both on classical and quantum level. However, it should be stressed that the sign of the integration constant cannot be fixed by group-theoretical considerations (cf. Ref. \cite{b18}) and the choice of the sign provides an additional input.

\par
To get rid of square root in action integral one can follow the standard procedure by writing
\begin{equation}
\label{e16}
\begin{split}
L=\sigma \left(2\gamma ^2\eta +\frac{1}{\eta }(\dot{z} +z^2)\right)  ,
\end{split}
\end{equation}
where $\eta $\ is the adjoint field transforming according to the eq. (\ref{e6}) with $d_\eta =-1$.\\
Note  that eq. (\ref{e16}) coincides with eq. (2.21 ) in Ref. \cite{b18} provided the identification $\eta =\exp {(x^3)}$, $ z=x^2$\ has been made.
\par
Now let us perform simple canonical analysis 
(cf. Ref. \cite{b19a}). 
The primary constraints read
\begin{equation}
\label{e17}
\begin{split}
\chi _1\equiv p_\eta\approx 0 , \quad \chi _2\equiv p_z-\frac{\sigma } {\eta }\approx 0 ,
\end{split}
\end{equation}
 while the Hamiltonian is written as 
 \begin{equation}
\label{e18}
\begin{split}
H=-2\sigma \gamma ^2\eta -\frac{\sigma }{\eta }z^2+u_\eta p_\eta +u_z(p_z-\frac{\sigma }{\eta }) ,
\end{split}
\end{equation} 
 $u_\eta , u_z$\ being the appropriate Lagrange multipliers. Imposing 
\begin{equation}
\label{e19}
\begin{split}
\dot{p_\eta} \approx 0, \quad \frac{d}{dt}(p_z-\frac{\sigma }{\eta })\approx 0   ,
\end{split}
\end{equation} 
we find no new constraints while
\begin{equation}
\label{e20}
\begin{split}
u_z=2\gamma^2 \eta^2-z^2 , \quad u_\eta  =-2z  ,
\end{split}
\end{equation} 
So the constraints (\ref{e17}) are of second kind. This allows us to eliminate $p_\eta $\ and $p_z$\ at the expense 
of introducing Dirac bracket 
\begin{equation}
\label{e21}
\begin{split}
\{\eta ,z\}_D=\frac{\eta ^2 }{\sigma }  ,
\end{split}
\end{equation} 
Take $\sigma =-\frac{1}{2}$; then
\begin{equation}
\label{e22}
\begin{split}
H=\gamma ^2\eta +\frac{z^2}{2\eta }, \quad \{z,\eta \}_D=2\eta ^2  .
\end{split}
\end{equation} 
Putting
\begin{equation}
\label{e23}
\begin{split}
\eta =\frac{1}{\rho ^2}, \quad z=\frac{p_\rho }{\rho }  ,
\end{split}
\end{equation} 
one arrives at the standard form of conformal mechanics
\begin{equation}
\label{e24}
\begin{split}
H=\frac{1}{2}p_\rho ^2+\frac{\gamma ^2}{\rho ^2}, \quad \{\rho ,p_\rho \}_D=1  .
\end{split}
\end{equation} 
Other symmetry generators read
\begin{equation}
\label{e25}
\begin{split}
D&=\frac{z}{2\eta }-\frac{1}{2}(2\gamma ^2\eta +\frac{z^2}{\eta })t, \\
K&=\frac{1}{2\eta }-\frac{z}{\eta }t+\frac{1}{2}(2\gamma ^2\eta +\frac{z^2}{\eta })t^2,
\end{split}
\end{equation} 
and acquire the standard form when expressed in terms of $\rho $\ and $p_\rho $\ variables.
\par
 
We conclude this section showing that one can use directly eq. (\ref{e11}) to derive the standard form of conformal mechanics. Indeed, the Lagrangian
\begin{equation}
L=\sigma\sqrt{\dot z +z^2}
\end{equation}
yields
\begin{equation}
p_z=\frac{\sigma}{2\sqrt{\dot z+z^2}}
\end{equation}
and
\begin{equation}
H=p_z\dot z-L=\frac{-\sigma^2}{4p_z}-p_zz^2
\end{equation}
Defining the canonical transformations
\begin{equation}
z=\frac{p_\rho^2}{\rho}, \quad p_z=\frac{-\rho^2}{2}
\end{equation}
one finds 
\begin{equation}
H=\frac{p_\rho^2}{2}+\frac{2\sigma^2}{\rho^2}
\end{equation}
which coincides with eq. (\ref{e24}) provided $2\sigma^2=\gamma^2$. It seems that we obtain the model with positive coupling. However, using eq. (\ref{e12}) one can  easily conclude that $\dot z +z^2$ is positive or negative if the integration constant $\gamma^2$ is positive or negative, respectively. This implies that the coefficient $\sigma$ should be real resp. imaginary to yield real value of the action (\ref{e11}).

\section{Dynamical realizations of conformal Galilei algebras} 
The $l$-conformal Galilei algebra (for simplicity we restrict ourselves to three dimensional case) is spanned
by $H,D,K$ together with $so(3)$\ generators $J_k$\ and $2l+1$ additional generators $\vec{C}^{(n)}, n=0,1,...2l$,\ obeying 
\begin{align}
\label{e26}
&[H,\vec{C}^{(n)}]=in \vec{C}^{(n-1)},\nonumber\\
&[K,\vec{C}^{(n)}]=i(n-2l) \vec{C}^{(n+1)},\nonumber\\
&[D,\vec{C}^{(n)}]=i(n-l) \vec{C}^{(n)},\\
&[J_i,C_k^{(n)}]=i\varepsilon _{ikm} C_m^{(n)}\nonumber
\end{align}
Consider the nonlinear action of $l$-conformal group defined by selecting the subgroup generated by $\vec{J}$\ and $D$. 
With such a choice we are not dealing with the symmetric decomposition. However, the generators $H,K$\ and $\vec{C}^{(n)}$\
span the linear representation under adjoint action of stability subgroup. Therefore our realization linearizes on it. In order
 to construct invariant dynamics it is sufficient to respect the invariance under rotations and dilatations.
 \par
 Let us choose the following parametrization of coset manifold 
\begin{equation}
\label{e27}
\begin{split}
w=e^{itH}e^{i\vec {x}^{(n)}\vec {C}^{(n)}}e^{izK} ;
\end{split}
\end{equation}
note the difference with respect to the parametrization used in \cite{b14},\cite{b15}.\\
The Cartan forms
\begin{equation}
\label{e28}
\begin{split}
w^{-1}dw=i(\omega _HH+\omega _DD+\omega _KK+\vec{\omega }^{(n)}\vec{C}^{(n)})  ,
\end{split}
\end{equation} 
are given by eqs.(\ref{e7}) together with 
\begin{equation}
\label{e29}
\begin{split}
\vec{\omega }^{(n)}=\sum_{p=0}^n\dbinom{2l-p}{2l-n}(-z)^{n-p}\left( d\vec{x}^{(p)}-(p+1)\vec{x}^{(p+1)}dt\right) .
\end{split}
\end{equation} 
The forms $\vec{\omega }^{(n)}$\ are vectors under $SO(3)$\ while under dilatations 
\begin{equation}
\label{e30}
\begin{split}
\vec{\omega }'^{(n)}=e^{(l-n)u}\vec{\omega }^{(n)} .
\end{split}
\end{equation} 
Define the covariant derivatives 
\begin{equation}
\label{e31}
\begin{split}
\nabla \vec{x}^{(n)}\equiv \frac{\vec{\omega }^{(n)}}{\omega _H} ,
\end{split}
\end{equation} 
with dilatation dimension $l-n-1$. Let $\vec{\lambda }^{(n)}$\ be the additional (adjoint) variables with dilatation dimension $n-l$.
Consider the following first order Lagrangian
\begin{equation}
\label{e32}
\begin{split}
L=-\gamma ^2\eta -\frac{1}{2\eta }(\dot{z}+z^2) +\sum_{n=0}^{2l} \vec{\lambda }^{(n)}\nabla \vec{x}^{(n)},
\end{split}
\end{equation} 
By the very construction it yields the invariant action functional. The equations of motion are of the form
\begin{align}
\label{e33}
&2\gamma ^2\eta ^2-(\dot{z}+z^2)=0,\nonumber \\
&\dot{\eta }+2z\eta =0,\nonumber \\
&\dot{\vec{x}}^{(n)}-(n+1)\vec{x}^{(n+1)}=0, \quad n=0,...,2l , \\
&\sum_{n=0}^{2l-p}\dbinom{2l-p}{n}\frac{d}{dt}\left((-z)^n\vec{\lambda }^{(n+p)}\right)+p\sum_{n=0}^{2l-p+1}
\dbinom{2l-p+1}{n}(-z)^n\vec{\lambda }^{(n+p-1)}=0, \nonumber
\end{align} 
We see that they  decouple. The first two describe conformal mechanics. Then there is a set of equations for
$\vec{x}^{(n)}$\ describing free higher-derivative system. Finally, once $z(t)$\ is determined one can solve last equations
for $\vec{\lambda }^{(n)}$; they do not impose any further constraints on $z$.
\par
Eqs. (\ref{e33}) are explicitly dilatation and rotation invariant; also the time translation invariance is obvious.\\
The action of special conformal transformations $\exp (icK)$\ reads
\begin{equation}
\label{e35}
\begin{split}
&t'=\frac{t}{1-ct}, \\
&z'=(1-ct)^2z+(1-ct)c,\\
&u'=-2\ln (1-ct), \\
&\vec{x}\;'^{(n)}=\sum_{p=0}^{n}\dbinom{2l-p}{2l-n}c^{n-p}(1-ct)^{n+p-2l}\vec{x}^{(p)}, \\
&\vec{\lambda }\;'^{(n)}=(1-ct)^{2(l-n)}\vec{\lambda }^{(n)}, \\
&\eta '=(1-ct)^2\eta 
\end{split}
\end{equation} 
 It is again not difficult to check that the eqs. (\ref{e33}) are invariant under the above transformations. We shall verify
 here the invariance of last eqs. (\ref{e33}). Let
\begin{equation}
\label{e36}
\vec{\mu }^{(p)}=\sum_{n=0}^{2l-p}\dbinom{2l-p}{n}\frac{d}{dt}\left((-z)^n\vec{\lambda }^{(n+p)}\right) 
\end{equation}
Last eq.(\ref{e33}) reads now
\begin{equation}
\label{e37}
\dot{\vec{\mu }}^{(p)}+p\vec{\mu }^{(p-1)}=0
\end{equation}
while the transformation rule for $\vec{\mu }^{(p)}$\ is
\begin{equation}
\label{e38}
\vec{\mu }\;'^{(p)}=(1-ct)^{2(l-p)}\sum_{k=0}^{2l-p}\dbinom{2l-p}{k}\left(\frac{-c}{1-ct}\right)^k\vec{\mu  }^{(p+k)}
\end{equation}
It is now straightforward to check the invariance of eqs. (\ref{e37}).\\
Finally, consider the action of $exp(i\vec{y}^{(n)}\vec{C}^{(n)})$:
\begin{equation}
\label{e39}
\begin{split}
&t'=t, \\
&z'=z,\\
&u'=0, \\
&\vec{x}\;'^{(n)}=\vec{x}^{(n)}+\sum_{k=n}^{2l}\dbinom{k}{n}t^{k-n}\vec{y}^{(k)}, \\
&\vec{\lambda }\;'^{(n)}=\vec{\lambda }^{(n)}, \\
&\eta '=\eta 
\end{split}
\end{equation}
The invariance of eqs.(\ref{e33}) is next to obvious.
\section{Hamiltonian formalism} 
Our Lagrangian, being of first order, provides an example of constrained system. Let us remind \cite{b20} that given 
a Lagrangian
\begin{equation}
\label{e40}
L=\sum_ia_i(q)\dot{q}_i+b(q)
\end{equation}
with
\begin{equation}
\label{e41}
\det \left[\omega _{ik} \right]\equiv \det\left[\frac{\partial a_k}{\partial q_i}-\frac{\partial a_i}{\partial q_k}\right]\not=0
\end{equation}
one finds that the full set of second kind constraints reads
\begin{equation}
\label{e42}
p_i-a_i(q)\approx 0
\end{equation}
The Hamiltonian dynamics is defined by
\begin{equation}
\label{e43}
H=-L\mid _{\dot{q}=0}=-b(q)
\end{equation}
\begin{equation}
\label{e44}
\{q_i,q_k\}_D=(\omega ^{-1})_{ik}
\end{equation}
In our case we obtain
\begin{equation}
\label{e45}
H=\gamma ^2\eta +\frac{z^2}{2\eta } + \sum_{n=0}^{2l}\vec{\lambda }^{(n)}\sum_{p=0}^n\dbinom{2l-p}{2l-n}(p+1)(-z)^{n-p}\vec{x}^{(p+1)}
\end{equation}
together with
\begin{equation}
\label{e46}
\begin{split}
\{x_a^{(n)}, \lambda ^{(m)}_b\}_D&=z^{n-m}\dbinom{2l-m}{2l-n}\delta _{ab} \\
\{z,\eta \}_D&=2\eta ^2 \\
\{\vec{\lambda }^{(k)},\eta \}_D&=2(2l-k)\eta ^2\vec{\lambda }^{(k+1)} 
\end{split}
\end{equation}
Again, it is straightforward although slightly tedious to check that eqs.(\ref{e45}), (\ref{e46}) yield correct dynamics.

\section{Oscillator-like parametrization} 
Let us compare the present approach with the one presented in Ref.\cite{b14}. To this end we use the identity
\begin{equation}
\label{e47}
e^{i\vec{x}^{(n)}\vec{C}^{(n)} }e^{izK}e^{iuD}=e^{izK}e^{iuD}e^{i\vec{x}\;'^{(n)}\vec{C}^{(n)} }
\end{equation}
\begin{equation}
\label{e48}
\vec{x}\;'^{(n)}=\sum_{m=o}^n\dbinom{2l-m}{2l-n}(-\dot{\rho })^{n-m}\rho ^{m+n-2l}\vec{x}^{(m)}
\end{equation}
where the substitutions $z=\dot{\rho }/\rho,\; e^u=\rho ^2$\ have been made. With the help of eqs.(\ref{e33}) one 
 derives the following equations of motion for new values (\ref{e48})
\begin{equation}
\label{e49}
\rho ^2\dot{\vec{x}}\;'^{(n)}=(n+1)\vec{x}\;'^{(n+1)}-(2l-n+1)\gamma ^2\vec{x}\;'^{(n-1)}
\end{equation}
which coincide with eqs. (12) in Ref. \cite{b14}.\\
Eqs. (\ref{e48}), together with time redefinition
\begin{equation}
\label{e50}
\tilde{t} \equiv \int\frac{dt}{\rho ^2(t)}
\end{equation}
provide a generalization of Niederer's transformation \cite{b16}. Indeed, they relate higher derivative free motion described
by eqs. (\ref{e33}) to the system of coupled harmonic oscillators.
\par
It is interesting to note that, with the new definition of time, eq.(\ref{e50}), one can put the oscillator system
 into unconstrained Hamiltonian form provided $l$\ is half-integer, i.e. $2l$\ is odd. To see this we define \cite{b21}
\begin{equation}
\label{e51}
\begin{split}
&\vec{q}_k=k!\vec{x}\;'^{(k)}, \quad k=0,....l-\frac{1}{2} \\
&\vec{p}_k=(-1)^{l-\frac{1}{2}-k}(2l-k)!m\vec{x}\;'^{(2l-k)},
\end{split}
\end{equation} 
$m$\ being an arbitrary nonzero parameter (the "mass"). We impose the standard Poisson structure
\begin{equation}
\label{e52}
\{q_{ka},p_{jb}\}=\delta _{kj}\delta _{ab}.
\end{equation}
Then the Hamiltonian yielding the equations of motion
\begin{equation}
\label{e53}
\frac{d\vec{x}\;'^{(n)}}{d\tilde{t}}=(n+1)\vec{x}\;'^{(n+1)} -(2l-n+1)\vec{x}\;'^{(n-1)},
\end{equation}
(we put here $\gamma ^2=1$\ for simplicity) reads
\begin{align}
\label{e54}
H&=\sum_{k=0}^{l-\frac{3}{2}}\vec{p}_k\vec{q}_{k+1}+\frac{1}{2m}\vec{p}_{l-\frac{1}{2}}^{\;2}
-\sum_{k=0}^{l-\frac{3}{2}}(k+1)(2l-k)\vec{p}_{k+1}\vec{q}_{k} \nonumber \\
 &+\frac{1}{2}m(l+\frac{1}{2})^2\vec{q}_{l-\frac{1}{2}}^{\;2}.
\end{align}
For the original $\vec{x}$\ variables and original time the same transformations  (\ref{e51}) give the canonical 
formalism invariant under the action of centrally extended conformal Galilei group. The relevant generators
 read \cite{b5}, \cite{b6}
\begin{equation}
\label{e55}
\begin{split}
H&= \sum_{k=0}^{l-\frac{3}{2}}\vec{p}_k\vec{q}_{k+1}+\frac{1}{2m}\vec{p}_{l-\frac{1}{2}}^{\;2}, \\
K&=\frac{1}{2}m(l+\frac{1}{2})^2\vec{q}_{l-\frac{1}{2}}^{\;2}- \sum_{k=0}^{l-\frac{3}{2}}(2l-k)(k+1)\vec{q}_k\vec{p}_{k+1}, \\
D&=\sum_{k=0}^{l-\frac{1}{2}}(l-k)\vec{p}_k\vec{q}_k, \\
\vec{J}&=\sum_{k=0}^{l-\frac{1}{2}}\vec{q}_k\times \vec{p}_k ,
\end{split}
\end{equation} 
while the generators $\vec{C}^{(k)}$\ are represented, up to multiplicative constants, by $\vec{q}_k$\ and $\vec{p}_{2l-k}$
(cf. Ref. \cite{b6}).
\section{Conclusions} 

Let us summarize our results. We used the method of nonlinear realizations \cite{b9,b10} to construt dynamical systems invariant under the action of $l$-conformal Galilei group for both integer and half integer variables of $l$. Our aim was to put emphasis on Lagrangian and Hamiltonian formulation. Therefore, instead of imposing invariant constraints on Cartan forms we enlarged the stability subgroup  (which allows us to abandom one constraint) and added new variables which, in  turn, allow us to construct  simple invariant Lagrangian in such a way that these new degrees of freedom do not enter the dynamics of original ones. The resulting dynamical equations of motion are described by eqs. (\ref{e33}). They can be summarized as follows. We define the variable $\vec x$ by 
\begin{equation}
\vec x ^{(n)}=\frac{1}{n!}\frac{d^n\vec x}{dt^n}
\end{equation}
while the conformal mode $\rho$ is given by eqs. (\ref{e23}). Then the eqs. (\ref{e33}) are equivalent to 
\begin{equation}
\label{e57}
\overset{..}{\rho}-\frac{\gamma^2}{\rho^3}=0,\quad \frac{d^{2l+1}\vec x} {dt^{2l+1}}=0
\end{equation}

Once these equations are solved one can find the dynamics  of the auxiliary variables $\vec \lambda ^(n)$ using the last equation (\ref{e33}).
\par 
The $l$-conformal  Galilei  group acts as the group of point transformations; for example, the proper conformal transformations read 
\begin{equation}
\begin{split}
&t'=\frac{t}{1-ct}\\
&\rho'=\frac{\rho}{1-ct}\\
&\vec{x}'=\frac{\vec x}{(1-ct)^{2l}}\\
&\vec\lambda'^{(n)}=(1-ct)^{2(l-n)}\vec\lambda^{(n)}
\end{split}
\end{equation}
The characteristic property of the eqs. (\ref{e57}) is that the $\rho$ and $\vec x$ variables decouple completely. We have achieved this by the appropriate choice of the subgroup, on which the action of $l$-conformal group linearizes (rotations and dilatations which seems to be rather natural assumption) and the specific parametriazation of the coset manifold (cf. eq. (\ref{e27})).
\par As it has been already mentioned in the Introduction the properties of $l$-conformal  Galilei group depend on whether $l$ is half integer or integer. In the former case it admits central extension. Then the application of the orbit method \cite{b6} yields basically unique picture of Hamiltonian dynamics invariant under the (transitive) action of the $l$-conformal Galilei group. 
It can be further shown \cite{b6}  that (neglecting spin variables obeying trivial dynamics) the configuration space consists  eventually of the conformal mode $\rho$  and the vector variable $\vec x $ obeying eqs. (\ref{e57}). 
Moreover, the (nonextended) $l$-conformal; Galilei group acts as the symmetry group of point transformations according to the same rules as desribed above.
\par We have shown that this description is universal in the sense that it works whether $l$ is half integer or integer and results from nonlinear realizations of $l$ conformal linearizing on rotations and dilations. The difference between the case of $l$ integer or half integer is that the latter admits, besides the Hamiltonian formalism presented here, the alternative one where no additional variables are necessary. Note that, when $\vec \lambda^{(n)}$ variables  are present, the group action is no longer transitive and phase space is not a coadjoint orbit.
\par Denote by $\mathcal{C}$ the abelian group generated by $\vec C^{(n)}$'s, $n=0,\ldots,2l$. It can be shown, using the same reasoning as in the third Ref. \cite{b6}, that the maximal symmetry of eqs. (\ref{e57})  is $(SL(2,\mR)\times GL(3,\mR))\uplus \mathcal{C}$
. For half integer $l$ and the variational formalism based on the orbit method \cite{b6} the Noether symmetries (i.e. those leaving the action functional invariant) correspond to l-conformal group, i.e.  $(SL(2,\mR)\times SO(3))\uplus \mathcal{C}$. Our guess is that the variational formalism presented here can be reformulated (by considering the nonlinear realizations of  $(SL(2,\mR)\times GL(3,\mR))\uplus \mathcal{C}$ as  to include all symmetries of eqs. (\ref{e57}) as Noetherian ones.
\par Finally, let us compare our results with those of Galajinsky and Masterov \cite{b14}. The explicit decoupling of the $\rho$ and $\vec x$ modes depends strongly on the choice of parametrization of coset space. The variables used by Galajinsky and Masterov are related to the ones we use by eq. (\ref{e48}). In terms of them the decoupling is not explicit. In fact, the dynamics is now given by eqs. (\ref{e49}). The matrix entering the right-hand side is fully diagonalizable with pairwise complex conjugated eigenvalues (together with one zero eigenvalue for integer $l$). Therefore, there exists linear combinations of $\vec x$'s which obey second order differential equation. On the other hand, these linear combinations form, together  with the conformal mode, a manifold invariant under the action of $l$ conformal group.
This is because the  rotation  generators, together with any set of linear combinations of $\vec C^{(n)}$'s, generate a subgroup which can serve as a stability subgroup in constructing nonlinear realizations. This can be explicitly  seen by using the parametrization considered in Ref. \cite{b14}. Abandoming the $SO(3)$ subgroup one can write the composition law as follows:
\begin{equation}
\label{e56}
(g_1,x_1)\cdot (g_2,x_2)=(g_1g_2,D^{-1}(g_2)x_1+x_2);
\end{equation}
here $g_i\in SL(2,\mR)$ and $x_i\in\mathcal {C}$. We used the additive convenction so that $\mathcal{C}$ may be identified with the linear space spanned by the generators $\vec C^{(n)}$, while $D(g)$ is the spin $l$ representation of $SL(2,\mR)$.
Identifying $(g,x),\; g\in SL(2,\mR),\;x\in  \mathcal C$\ with elements of coset space we find the action of $l$-conformal Galilei group on the coset space   
\begin{equation}
\label{e56}
(g_1,a)\cdot (g,x)=(g_1g,D^{-1}(g)a+x),
\end{equation}
It is now obvious  that the submanifold defined by choosing any quotient space in $\mathcal{C}$  carries the nonlinear action of $l$-conformal Galilei group.

 Such a construction is not possible in our parametrization: the relevant matrix defining dynamical equations for $x's$\ is nilpotent (cf. eqs.  (\ref{e33})).
 Obviously, one can repeat the reasoning of Galajinsky and Masterov but the relevant combination of $x$\ variables has then $\rho $-dependent coefficients. 
 So we have either complete decoupling or invariant dynamics in terms of second order differential equations. The main problem in the latter case is 
 whether they admit Lagrangian and Hamiltonian description.

The construction of invariant Lagrangian is not straightforward because the group action does not linearize on the stability subgroup.

\par
{\bf Acknowledgments} The authors are grateful to Professor Jerzy Lukierski for useful remarks and discussion.
One of us (P.K.) thanks Professor Anton Galajinsky for kind correspondence and helpful remarks. 
The remarks of anonymous referee which helped to improve the paper are gratefully acknowledged.
This work  is supported  in part by  MNiSzW grant No. N202331139

\end{document}